\documentstyle[12pt]{article}

\begin{document}

\title{Masses of spin-1 mesons and dynamical chiral symmetry breaking}
\author{Bing An Li\\Department of Physics and Astronomy, University
of Kentucky\\Lexington, KY 40506, USA}
\maketitle

\begin{abstract}
In terms of an effective chiral theory of mesons
it is shown quantitatively
that in the limit of $m_{q}=0$, the masses of $\rho$, $\omega$,
$K^{*}(892)$, $m_{\phi}$,
$a_{1}$, $f_{1}(1286)$, $K_{1}(1400)$, and $f_{1}(1510)$
mesons originate from dynamical chiral symmetry breaking.
\end{abstract}
\newpage

The origin of the masses of hadrons is always one of the most
important topics in hadron dynamics. The mass of a hadron is
associated with chiral symmetry breaking. There are implicit chiral
symmetry breaking from the current quark masses
and dynamical chiral symmetry breaking. The Nambu-Jona-Lasinio
model[1], starting from massless quark, has nonvanishing quark
condensate which means dynamical chiral symmetry breaking. On the
other hand, dynamical chiral symmetry breaking in Quantum
Chromodynamics($QCD$) has been studied extensively[2].
$QCD$ has dynamical chiral symmetry breaking.
It is well known that
in the chiral limit, the octet pseudoscalar
mesons are Goldstone mesons and massless.
{}From the theory
of chiral symmetry breaking Gell-Mann, Oakes, and Renner[3]
have obtained
\begin{equation}
m^{2}_{\pi}=-{2\over f^{2}_{\pi}}(m_{u}+m_{d})<0|\bar{\psi}\psi|0>.
\end{equation}
The smallness of the masses of the current quarks leads to the light
pseudoscalar mesons.
The heavy $\eta'$ meson
is associated with
$U(1)$ problem[4]. Why the mass of the $\rho$ meson is much heavier
than pion's mass? In this letter we try to find the origin of the
$\rho$ meson's mass.
In Ref.[5] we have proposed an effective chiral
theory of mesons(pions, $\eta$, $\rho$, $\omega$, $a_{1}(1260)$, and
$f_{1}(1286)$).
In the case of two flavors
the Lagrangian of this theory is constructed by
using $U(2)_{L}\times U(2)_{R}$ chiral symmetry and the minimum
coupling principle
\begin{eqnarray}
{\cal L}=\bar{\psi}(x)(i\gamma\cdot\partial+\gamma\cdot v
+\gamma\cdot a\gamma_{5}
-mu(x))\psi(x)\nonumber \\
+{1\over 2}m^{2}_{0}(\rho^{\mu}_{i}\rho_{\mu i}+
\omega^{\mu}\omega_{\mu}+a^{\mu}_{i}a_{\mu i}+f^{\mu}f_{\mu})
-\bar{\psi}(x)M\psi(x),
\end{eqnarray}
where \(a_{\mu}=\tau_{i}a^{i}_{\mu}+f_{\mu}\), \(v_{\mu}=\tau_{i}
\rho^{i}_{\mu}+\omega_{\mu}\),
and \(u=exp\{i\gamma_{5}(\tau_{i}\pi_{i}+
\eta)\}\), $m$ is a parameter and M is the quark mass matrix.
In Eq.(2) $u$ can be written as
\begin{equation}
u={1\over 2}(1+\gamma_{5})U+{1\over 2}(1+\gamma_{5})U^{\dag},
\end{equation}
where \(U=exp\{i(\tau_{i}\pi_{i}+\eta)\}\).
The introduction of the couplings
between the pseudoscalar mesons and the quarks is based on the
formalism of the nonlinear $\sigma$ model.
The Vector Meson Dominance(VMD) is a natural result of this theory,
Weinberg's first sum rule is derived from this theory analytically.
A unified study of the processes of normal parity and abnormal parity
is presented by this theory. Most
theoretical results agree with the data within about $10\%$.
Besides the success in the
phenomenology of mesons, this theory has dynamical chiral symmetry
breaking(Eq.(124) of Ref.[5])
\begin{equation}
<0|\bar{\psi}(x)\psi(x)|0>=3m^{3}g^{2}(1+{1\over 2\pi^{2}g^{2}}),
\end{equation}
where g is the universal
coupling constant and defined as
\begin{equation}
g^{2}={f^{2}_{\pi}\over m^{2}_{\rho}}+{f^{2}_{\pi}\over 6m^{2}},
\end{equation}
where $f_{\pi}$ is the pion decay constant and \(f_{\pi}=186MeV\),
\(m^{2}_{\rho}=m^{2}_{0}/g^{2}\).
Eq.(4) means that the parameter m of the Lagrangian(Eq.(2)) is the
indication of the dynamical chiral symmetry breaking.
Use of the Eq.(4) leads to the mass formula of pion(Eq.(1))[5].
In the chiral limit, there are three parameters: cutoff $\Lambda$,
m, and
$m_{\rho}$ and g is determined by the ratio $\Lambda/m$(see Eq.(129)
of Ref.[5]).
The purpose of this letter is to illustrate
that in the limit of \(m_{q}=0\), the masses of
spin-one mesons are resulted by dynamical chiral symmetry breaking
and $m_{\rho}$ can be determined
theoretically and is no longer an input.
The introduction of the mass terms of the spin-1 mesons to the
Lagrangian(Eq.(2)) is necessary. Otherwise, this theory cannot be well
defined. For example, without these mass terms the mixing between
the axial-vector meson and the corresponding pseudoscalar meson
cannot be erased, the physical axial-vector and pseudoscalar fields
cannot be defined. On the other hand, in the effective Lagrangian of
meson fields derived from Eq.(2), except for the kinetic terms of the
spin-1 mesons, the spin-1 fields only appear in the covariant
derivatives(see Eq.(13) of Ref.[5])
\[D_{\mu}U=\partial_{\mu}U-i[v_{\mu}, U]+i\{a_{\mu}, U\}.\]
Because the $a_{\mu}$ fields are in the anticommutator, a spontaneous
symmetry breaking mechanism leads to the mass difference between
the axial-vector and vector mesons. However, the vector
fields are in the commutator, there is no way to generate a mass
term for the vector meson in this theory.
Therefore, the mass terms of spin-1 mesons
must be introduced.

In $QCD$, in principle, the mass of the $\rho$ meson should be
determined by a dynamical equation of bound state. In this theory
KSFR sum rule[6]
\begin{equation}
g_{\rho}={1\over 2}f_{\rho\pi\pi}f^{2}_{\pi}
\end{equation}
can be taken as the equation used to determine
$m_{\rho}$. This sum rule
is obtained by using PCAC and current algebra in the limit of
$p_{\pi}\rightarrow 0$. $g_{\rho}$ is the coupling constant
of $\rho-\gamma$ and $f_{\rho\pi\pi}$ is the coupling constant
of $\rho\rightarrow \pi\pi$, which is
defined in the limit of $p_{\pi}\rightarrow 0$.
In Ref.[5] KSFR sum rule is satisfied numerically. On the other hand,
this sum rule can be derived analytically from this theory.
{}From Eq.(2) the currents are found to be
\[V^{i}_{\mu}=\bar{\psi}{\tau_{i}\over 2}\gamma_{\mu}\psi,\;\;\;
A^{i}_{\mu}=
\bar{\psi}{\tau_{i}\over 2}\gamma_{\mu}\gamma_{5}\psi.\]
These currents observe current algebra.
Without the quark mass term the Lagrangian(Eq.(2)) is global
$U(2)_{L}\times U(2)_{R}$ chiral
symmetric and both the isovector axial-vector current are conserved.
The mass term of the pion is generated from the
quark mass term[7]
\begin{equation}
-\bar{\psi}(x)M\psi(x)=iTrMs_{F}(x,x)=\frac{i}{(2\pi)^{D}}TrM
\int d^{D}ps_{F}(x,p),
\end{equation}
where the quark propagator satisfies the equation[5]
\begin{equation}
\{\gamma\cdot(i\partial+p+v+a\gamma_{5})-mu-M\}s_{F}(x,p)=1.
\end{equation}
Using the solution of Eq.(8)[5] and
to the leading order in quark mass expansion, we obtain
\begin{equation}
-M\bar{\psi}M\psi=-{1\over2}m^{2}_{\pi}\pi_{i}\pi_{i}+\cdots,
\end{equation}
where $m^{2}_{\pi}$ is expressed by Eq.(1).
With the pion mass term PCAC is derived from
the Lagrangian(Eq.(2))
\[\partial^{\mu}A^{i}_{\mu}={1\over2}m^{2}_{\pi}f_{\pi}\pi^{i}.\]
Therefore, the KSFR sum rule can be
derived from this theory in the same way as in Ref.[6].
It is necessary to point out that
due to the limit of $p_{\pi}\rightarrow 0$,
the coupling constant
$f_{\rho\pi\pi}$ of Eq.(6) is defined in nonphysical region.
However, KSFR sum rule agrees with data within $10\%$. This fact
means that in the limit of $p_{\pi}\rightarrow 0$,
$f_{\rho\pi\pi}$ is very close to the physical one.
The effective chiral theory[5] provides an explanation to this
property of $f_{\rho\pi\pi}$.
The expression of $f_{\rho\pi\pi}$ determined in Ref.[5]
(see Eq.(48) of [5]) shows a weak dependence of $f_{\rho\pi\pi}$
on the pion momentum
(in the reasonable range of the coupling constant g,
the contribution of the terms which are proportional to the pion
momentum is less than about $10\%$). In Ref.[5] \(g=0.35\)
is chosen and
\begin{equation}
f_{\rho\pi\pi}={2\over g},
\end{equation}
which is independent of $p_{\pi}$. $g_{\rho}$ is
determined to be[5]
\begin{equation}
g_{\rho}={1\over 2}gm^{2}_{\rho}.
\end{equation}
Substituting Eqs.(10),(11) into the KSFR sum rule it is found
\begin{equation}
m^{2}_{\rho}=2{f^{2}_{\pi}\over g^{2}}.
\end{equation}
Using Eq.(5), we obtain
\begin{equation}
m^{2}_{\rho}=6m^{2}.
\end{equation}
Therefore, the mass of $\rho$ meson is no longer an input and
is determined by the parameter m of Eq.(2) completely. In the
limit of \(m_{q}=0\), there are only two parameters
in the Lagrangian(Eq.(2)), which are cutoff $\Lambda$ and m.
Using Eq.(4), we obtain
\begin{equation}
m^{2}_{\rho}=6<0|\bar{\psi}\psi|0>^{{2\over 3}}(3g^{2}+
{3\over 2\pi^{2}})^{-{2\over 3}}.
\end{equation}
In the limit of \(m_{q}=0\),
the mass of $\rho$ meson originates
from dynamical chiral symmetry breaking. Combining Eqs.(5),(13),
we obtain
\begin{equation}
m^{2}=\frac{f^{2}_{\pi}}{3g^{2}}=0.094GeV^{2},\;\;\;
<0|\bar{\psi}\psi|0>=-(0.247)^{3}GeV^{3}.
\end{equation}
The mass of $\rho$ meson is determined to be
\begin{equation}
m_{\rho}=0.751GeV.
\end{equation}
It is only $2\%$ away from the experimental value of 0.77GeV.
The numerical value of m is slightly
different from the one presented in Ref.[5]. The reason is that
in Ref.[5] the physical value of $m_{\rho}$ is taken as input.
In the limit of \(m_{q}=0\)(\(q=u, d, s\)), we should have
\begin{equation}
m_{\phi}=m_{K^{*}(892)}=m_{\omega}=m_{\rho}.
\end{equation}
Therefore, in the limit of \(m_{q}=0\),
the masses of the four low lying vector mesons originate from
dynamical chiral symmetry breaking.
{}From Eqs.(4),(12),(13) it is found that
\begin{equation}
f^{2}_{\pi}=3g^{2}m^{2}=3g^{2}
<0|\bar{\psi}\psi|0>^{{2\over 3}}(3g^{2}+
{3\over 2\pi^{2}})^{-{2\over 3}}.
\end{equation}
The pion decay constant is the result of
dynamical chiral symmetry breaking too.
Therefore, in the limit of $m_{q}\rightarrow 0$,
the decay constants of the octet pseudoscalar mesons originate
from dynamical chiral symmetry breaking.
Using Eqs.(1),(14),and (18) we obtain
\begin{equation}
\frac{m^{2}_{\pi}}{m^{2}_{\rho}}=-\frac{(g^{2}+{1\over2\pi^{2}})^
{{4\over3}}}{3^{{2\over3}}g^{2}}\frac{m_{u}+m_{d}}
{<0|\bar{\psi}\psi|0>^{{1\over3}}}.
\end{equation}

In Ref.[5] the mass relations have been found
\begin{eqnarray}
(1-{1\over 2\pi^{2}g^{2}})m^{2}_{a}
=6m^{2}+m^{2}_{\rho},\;\;\;
(1-{1\over 2\pi^{2}g^{2}})m^{2}_{f}=
=6m^{2}+m^{2}_{\omega},
\end{eqnarray}
where $m_{a}$ is the mass of the meson $a_{1}(1260)$ and $m_{f}$
is the mass of the meson $f_{1}(1286)$.
In this theory \(m_{\rho}=m_{\omega}\), the theory predicts \(
m_{f}=m_{a}\). Using Eq.(13), the mass relations(Eq.(20)) are rewritten
as
\begin{equation}
m^{2}_{f}=m^{2}_{a}=2m^{2}_{\rho}{g^{2}_{a}\over g^{2}_{\rho}},
\end{equation}
where \(g_{a}={1\over 2}m^{2}_{\rho}g^{2}(1-{1\over 2\pi^{2}g^{2}})
^{-{1\over 2}}\) determined in Ref.[5].
This relation is different from Weinberg's second sum rule[8], as
pointed in Ref.[9], it is the result of Weinberg's first sum rule[8]
and KSFR sum rule.
On the other hand, the relation between $m^{2}_{a}$ and the quark
condensate is established by using Eq.(14)
\begin{equation}
m^{2}_{f}=m^{2}_{a}=12{g^{2}_{a}\over g^{2}_{\rho}}
<0|\bar{\psi}\psi|0>^{{2\over 3}}(3g^{2}+
{3\over 2\pi^{2}})^{-{2\over 3}}.
\end{equation}
In the limit of \(m_{q}=0\),
the masses of $a_{1}(1260)$ and
$f_{1}(1286)$ are resulted in dynamical chiral symmetry breaking.
Using \(g=0.35\), we obtain \(m_{f}=m_{a}=1388 MeV\) and the
experimental data are \(m_{a}=1230\pm 40MeV\) and \(m_{f}=1282\pm
2 MeV\). The deviations are about $10\%$.

The mass relations have been
obtained in Ref.[10] in which the theory is generalized to include the
strange quark
\begin{eqnarray}
(1-{1\over 2\pi^{2}g^{2}})m^{2}_{K_{1}(1400)}
=6m^{2}+m^{2}_{K^{*}(892)},\;\;\;
(1-{1\over 2\pi^{2}g^{2}})m^{2}_{f_{1}(1510)}=6m^{2}
+m^{2}_{\phi}.
\end{eqnarray}
Use of Eq.(13) leads to
\begin{eqnarray}
m^{2}_{K_{1}(1400)}
={g^{2}_{a}\over g^{2}_{\rho}}(m^{2}_{\rho}+m^{2}_{K^{*}(892)}),\;\;\;
m^{2}_{f_{1}(1510)}={g^{2}_{a}\over g^{2}_{\rho}}(m^{2}_{\rho}
+m^{2}_{\phi}).
\end{eqnarray}
Substituting Eq.(21) into Eq.(24) we obtain
\begin{equation}
m^{2}_{K_{1}(1400)}=\frac{m^{2}_{a}}{2}(1+{m^{2}_{K^{*}(892)}\over
m^{2}_{\rho}}),\;\;\;
m^{2}_{f_{1}(1510)}=\frac{m^{2}_{a}}{2}(1+{m^{2}_{\phi}\over
m^{2}_{\rho}}).
\end{equation}
The deviations of these mass relations from the experimental values are
about $3\%$.
By using Eq.(14), the dependences of $m^{2}_{K_{1}(1400)}$ and
$m^{2}_{f_{1}(1510)}$ on the quark condensate are obtained
\begin{eqnarray}
m^{2}_{K_{1}(1400)}=12{g^{2}_{a}\over g^{2}_{\rho}}
<0|\bar{\psi}\psi|0>^{{2\over 3}}(3g^{2}+
{3\over 2\pi^{2}})^{-{2\over 3}}+{g^{2}_{a}\over g^{2}_{\rho}}
(m^{2}_{K^{*}(892)}-m^{2}_{\rho})\nonumber \\
m^{2}_{f_{1}(1510)}=12{g^{2}_{a}\over g^{2}_{\rho}}
<0|\bar{\psi}\psi|0>^{{2\over 3}}(3g^{2}+
{3\over 2\pi^{2}})^{-{2\over 3}}+{g^{2}_{a}\over g^{2}_{\rho}}
(m^{2}_{\phi}-m^{2}_{\rho}).
\end{eqnarray}
It is well known that the mass differences of $m^{2}_{K^{*}(892)}-
m^{2}_{\rho}$ and $m^{2}_{\phi}-m^{2}_{\rho}$ are proportional to
the masses of quarks. From Eqs.(22),(26) we conclude that
in the limit of \(m_{q}=0\), the masses
of the four axial-vector mesons originate from dynamical
chiral symmetry breaking.

To conclude, in the limit of \(m_{q}\rightarrow 0\),
in the effective chiral theory the masses of the eight
spin-one mesons and
the decay constants of the octet pseudoscalar mesons originate from
dynamical chiral symmetry breaking. Three new mass relations(Eqs.(21),
(24)) are found.

This research is partially
supported by DOE grant DE-91ER75661.


\begin{thebibliography}{10}
\bibitem Y.Nambu and G.Jona-Lasinio, Phys. Rev. {\bf 122},345(1961);
{\bf 124},246(1961). R.Finger and J.E.Mondula, Nucl. Phys.
{\bf B199}, 168(1982).
\bibitem{} K.Lane, Phys. Rev. {\bf D100}, 2605(1974); H.D.Politzer,
Nucl. Phys. {\bf B117}, 397(1976); H.Pagels, Phys. Rev. {\bf D19},
3080(1979); V.Elias and M.D.Scadron, Phys. Rev. {\bf D30},
647(1984); L.N.Chang and N.P.Chang, Phys. Rev. {\bf D29}, 312
(1984); G.Krein, P.Tang, and A.G.Williams, Phys. Lett. {\bf B215},
145(1988); J.Kogut et al., Phys. Rev. Lett. {\bf 48}, 1140(1982).
\bibitem{} M.Gell-mann, R.J.Oakes, and B.Renner, Phys.Rev., {\bf 175},
195(1968).
\bibitem{} E. Witten, Nucl. Phys., {\bf B149}, 285(1979); G. Veneziano,
Nucl. Phys., {\bf B159},213(1979); C. Rosenzweig, J. Schechter, and
C. G. Trahern, Phys. Rev., {\bf D21},3388(1980); P. Nath and
R. Arnowitz,
Phys. Rev., {\bf D23},473(1981).
\bibitem{} Bing An Li, $U(2)_{L}\times U(2)_{R}$ Chiral Theory of
Mesons, to appear in {\bf 52D} No.9(1995).
\bibitem{} K.Kawarabayashi and M.Suzuki, Phys.Rev.Lett., {\bf 16}
255,(1966); Riazudin and Fayyazudin, Phys.Rev., {\bf 147},1071(1966).
\bibitem{} Bing An LI, Phys.Rev., {\bf D50},2243(1994).
\bibitem{} S.Weinberg, Phys.Rev.Lett., {\bf 17}, 616(1966).
\bibitem{} Bing An Li, Talk presented at the International Europhysics
Conference on High Energy Physics(HEP95),
July 27-Aug.2, Brussels, Belgium.
\bibitem{} Bing An Li, $U(3)_{L}\times U(3)_{R}$ Chiral Theory of
Mesons, to appear in {\bf 52D}, No.9(1995).


\end{thebibliography}
\end{document}